\documentclass[jkps,fleqn,showkeys]{revtex4}
\usepackage[pdftex]{graphicx}
\usepackage{amssymb}
\usepackage{amsmath}
\usepackage{bm}
\usepackage{xcolor}
\definecolor{dkgreen}{rgb}{0,0.6,0}
\definecolor{blue-violet}{rgb}{0.54, 0.17, 0.89}
\usepackage{kotex}

\begin{document}
\setcounter{page}{1}
\title[]{Effects of spatial noise on absorbing phase transitions in evolutionary cooperation dynamics}
\title[]{Effects of spatial environmental noise on evolution of cooperation}

\author{Janguk \surname{Kim}}
\affiliation{Department of Physics, Fudan University, Shanghai, 200437, People's Republic of China}
\author{Seung-Woo \surname{Son}}
\email{sonswoo@hanyang.ac.kr}
\affiliation{Department of Applied Physics, Hanyang University ERICA, Ansan, 15588, Republic of Korea}
\author{Hye Jin \surname{Park}}
\email{hyejin.park@inha.ac.kr}
\affiliation{Department of Physics, Inha University, Incheon, 22212, Republic of Korea}


\begin{abstract}
We investigate the effects of environmental noise on cooperation in a spatial evolutionary game model with variable population size. Building on a one-dimensional lattice model in which vacancies promote cooperation through spatial selection, we add random noise to the environmental quality parameter and consider two distinct types: annealed noise, where the environmental quality fluctuates independently at each site and each time step, and quenched noise, where each site is assigned a permanently fixed random value. For annealed noise, we develop a mean-field theory by replacing the noise-dependent death probabilities with their distribution averages, and find that increasing the noise intensity shifts both the cooperator-defector phase boundary and the absorbing boundary upward in the parameter space, simultaneously expanding the cooperative regime and the extinction region. These predictions are confirmed by numerical simulations. In contrast, quenched noise leaves the phase boundary nearly unchanged across all noise levels, exerting only a weak effect on cooperator frequency. Together, these results demonstrate that temporal fluctuations, rather than static spatial heterogeneity, are the primary driver of noise-induced shifts in the cooperative phase structure.
\end{abstract}



\maketitle

\section{INTRODUCTION}
Understanding how cooperation can arise and persist in social dilemma settings remains a central challenge in evolutionary game theory\cite{R2, R10, R11, R12, R20}. In the framework of social dilemmas, cooperators provide a collective benefit at a personal cost, while defectors exploit this benefit without contributing. As a result, defectors enjoy a short-term fitness advantage, which under standard well-mixed population dynamics drives cooperators toward extinction. Spatial structure has long been recognized as a key mechanism capable of reversing this outcome. When interactions are confined to local neighborhoods, cooperators can form compact clusters, thereby shielding themselves from exploitation by defectors and sustaining cooperation through what is known as `network' or `spatial' reciprocity\cite{R3, R4, R13, R14, R21, R22, R23, R24, R25, R30}. 

More recently, attention has turned to frameworks in which evolutionary strategy dynamics are coupled with population density dynamics, so that the composition and size of the population co-evolve\cite{R16, R17, R27, R28}. Such eco-evolutionary models with spatial dynamics reveal that feedback between ecological and evolutionary processes can qualitatively alter the fate of cooperation, supporting its emergence and maintenance under conditions where purely frequency-based dynamics would predict its collapse. In one-dimensional systems in particular, the competition between cooperative and defective domains can drive the population through absorbing phase transitions into extinction, whose critical behavior has been connected to the directed percolation universality class\cite{R9, R15, R26}.

In particular, Park~et~al.\cite{R1} showed that naturally occurring vacancies in growing one-dimensional habitats strongly promote cooperation: empty sites buffer cooperators from defectors, allowing them to survive in harsh environments where defectors go extinct. However, the quality of the environment was treated as a fixed constant in that framework. Environmental stochasticity, arising from temporal fluctuations or persistent spatial heterogeneity, is a ubiquitous feature of real ecological systems that can fundamentally alter the course of evolutionary dynamics\cite{R4, R5, R6, R7, R8, R18, R19, R29}. Yet the effect of environmental stochasticity in spatial games with variable population size remains unexplored.

In this paper, we build on the framework of Park~et~al.\cite{R1} and investigate how annealed and quenched environmental noise affect the cooperator-defector (C-D) phase boundary in the growing-habitat model. In the annealed case, the environmental parameter fluctuates independently at each site and each time step. In the quenched case, each site retains a fixed random value throughout the simulation, representing permanently heterogeneous spatial conditions~\cite{R5}. These two types of disorder have qualitatively different physical origins and are expected to produce distinct effects on cooperative dynamics\cite{R8, R9}. For annealed noise, we develop a mean-field (MF) theory and show analytically that increasing the noise intensity shifts the phase boundaries. For quenched noise, treated numerically, the C-D phase boundary remains unchanged. 

\section{Model}

\begin{figure}[t!]
\includegraphics[width=0.9\linewidth]{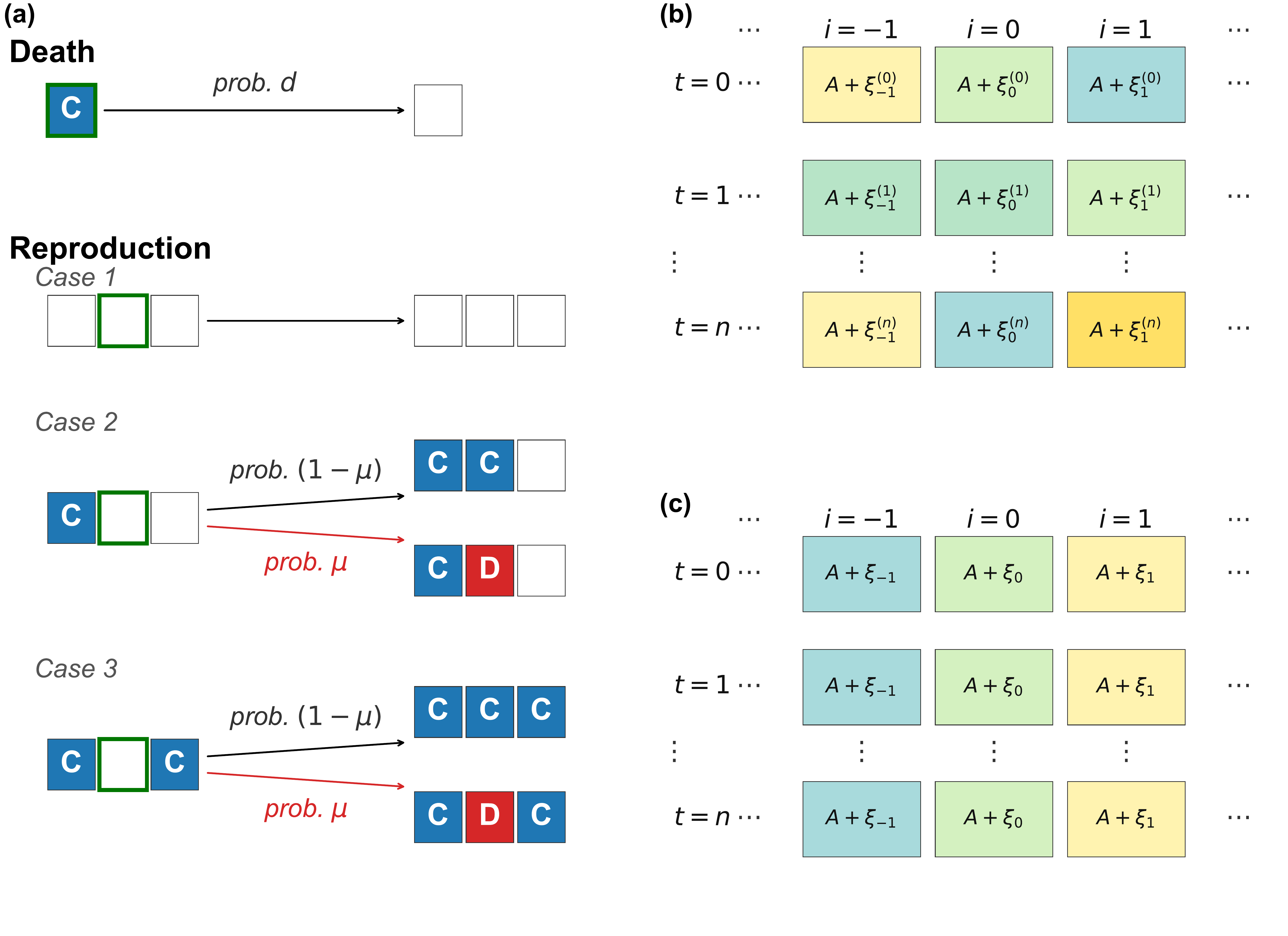}
\caption{
Schematic figure describing (a) death and reproduction processes and (b, c) annealed and quenched spatial noise on environmental quality. (a) Model dynamics on a 1D lattice. At each step, a site is chosen at random (green outline). If occupied, the individual dies with probability $d$ and the site becomes vacant (death process). If empty, a neighboring individual reproduces into it (reproduction process): `Case 1' shows no occupied neighbor case; `Case 2' shows reproduction when only one of two neighbors is occupied, with the offspring inheriting the parent's strategy with probability $1-\mu$ or mutating with probability $\mu$; `Case 3' shows reproduction when both neighbors are occupied, in which case either parent is chosen with equal probability. Empty, blue, and red boxes represent vacant sites (E), cooperators (C), and defectors (D), respectively. (b) Annealed spatial noise: at each MC step, an independent noise value $\xi_i \sim \mathcal{N}(0,\sigma^2)$ is drawn for every site, so the effective environmental parameter $A+\xi_i$ varies independently in both space and time. (c) Quenched spatial noise: each site is assigned a fixed noise value $\xi_i$ drawn once at the start of the simulation and held constant throughout, representing a permanently heterogeneous spatial environment.}
\label{fig1}
\end{figure}

We consider a one-dimensional infinite lattice model in which each site can be empty (E) or occupied by either a cooperator (C) or a defector (D), as shown in Fig.1(a). Players interact with each of their neighbors located at nearest neighboring sites via a donation game, in which cooperators pay a cost $c$ to provide a benefit $b$ to each neighbor, while defectors pay no cost and provide no benefit\cite{R1}. Normalizing $b=1$ and restricting to the social dilemma case $0<c<1$~\cite{R2, R3}, the payoff matrix is given by
\begin{equation}
    \begin{pmatrix} 1-c & -c & 0 \\ 1 & 0 & 0 \end{pmatrix},
    \label{payoff}
\end{equation}
where rows correspond to the focal player (C or D) and columns to the type of neighbor's site (C, D or E). The total payoff $p$ of each individual is the sum of payoffs from interactions, and is fully determined by the types of its two immediate neighbors according to the payoff matrix written in Eq.~(\ref{payoff}). 

We assume that a player's death probability is determined by its payoff $p$ and the quality of the environment factor $A\geq 0$~\cite{R1},
\begin{equation}
    d = \frac{1}{1+Ae^{wp}}.
    \label{deathprob}
\end{equation}
The selection strength $w\ge0$ controls how strongly the death probability depends on payoff. Larger $w$ indicates a stronger payoff-dependence of the death probability. We set $w=1$ throughout our study, as our focus is on the heterogeneity of $A$. A larger value of $A$ corresponds to a more favorable environment in which individuals are less likely to die, while the death probability increases for a smaller value of $A$. In the limit $A\rightarrow0$, death is certain regardless of payoff, whereas for sufficiently large $A$ the death rate approaches zero. 

The habitat is defined as the region between the leftmost occupied site $L$ and the rightmost occupied site $R$, with size $H(t)=R(t)-L(t)+1$ at time $t$\cite{R1}. One Monte Carlo (MC) step consists of $H+2$ elementary update trials. In each trial, a site is chosen uniformly at random from the $H+2$ sites spanning $[L-1,R+1]$. If the chosen site is occupied, the individual dies with probability $d$ given by Eq.~(\ref{deathprob}) and the site becomes vacant. If the chosen site is empty, a neighboring individual located at the adjacent site reproduces into it. When only one adjacent neighbor is occupied, its offspring fills the site, while if both adjacent neighbors are occupied, either is chosen with equal probability. During reproduction, the offspring inherits the parent's strategy with probability $1-\mu$ and mutates to the opposite strategy with probability $\mu$, as illustrated in Fig.~\ref{fig1}(a). Because the habitat boundaries advance when boundary-adjacent empty sites are colonized and retreat when individuals at the boundary die, the habitat size $H(t)$ is a dynamical variable that can grow, shrink, or go extinct\cite{R1, R3}.

To incorporate spatial environmental noise, we assign to each site $i$ an environmental quality $A_i=A+\xi_i$, where $\xi_i \sim \mathcal{N}(0, \sigma^2)$ is drawn independently for each site. The mean environmental quality is thus $A$, with fluctuations of magnitude $\sigma$. We consider both annealed and quenched noise, illustrated in Figs.~\ref{fig1}(b) and \ref{fig1}(c). In the annealed case, $\xi_i$ is resampled at every MC step, producing spatiotemporal fluctuations in environmental quality\cite{R4}; in the quenched case, $\xi_i$ is fixed throughout the simulation, representing a permanently heterogeneous spatial environment such as persistent differences in local habitat quality\cite{R5}. In both cases, $A_i$ is bounded below by $\epsilon = 0.01$ to prevent unphysical negative values. The results are robust even when a smaller value is chosen for the lower bound.

Starting from a single cooperator (seed individual), we simulate the habitat dynamics and calculate the cooperator frequency $x_C$, the fraction of cooperators among all living individuals, once the system reaches steady state. Each data point is averaged over non-extinct realizations among $M$ independent runs (ensembles). Although we used a cooperator as an initial seed, the choice of seed type does not affect the results at steady state in the presence of mutations.
To examine how spatial environmental noise shapes the phase structure, we find the cooperator-dominant and defector-dominant phases in $A$-$c$ space across a range of noise levels $\sigma$, for both annealed and quenched disorder. For the annealed case, we further derive the phase boundary analytically and verify that it is consistent with our numerical results.

\section{Results}
\begin{figure}[t!]
\includegraphics[width=\linewidth]{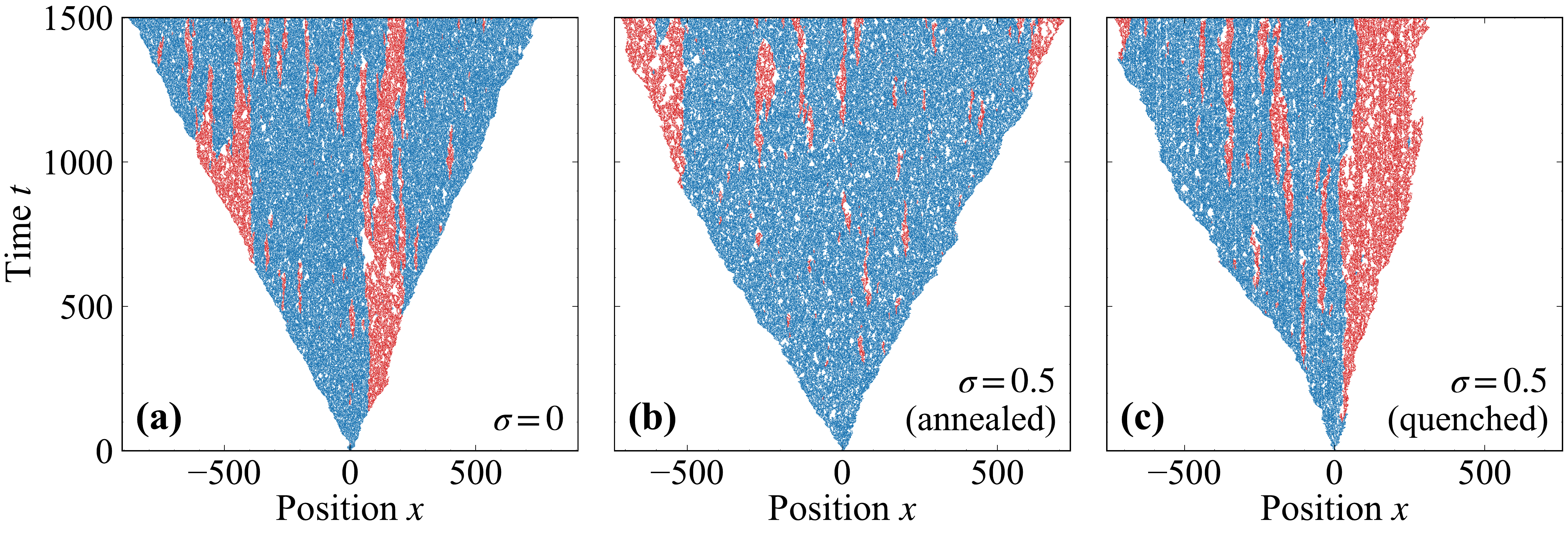}
\caption{Space-time plots illustrating the effect of spatial environmental noise on population dynamics, for $A=0.9$, $c=0.6$, $w=1$, and $\mu = 0.01$. Blue, red, and white indicate cooperators (C), defectors (D), and vacant sites (E), respectively; all populations are seeded from a single cooperator. (a) $\sigma = 0$ (no noise): cooperators and defectors coexist and form well-defined spatial domains. (b) $\sigma = 0.5$ (annealed noise): the cooperator-occupied region visibly expands relative to the noiseless case, reflecting the promotion of cooperation by temporal environmental fluctuations. (c) $\sigma = 0.5$ (quenched noise): the frozen spatial heterogeneity creates sites with sufficiently low environmental quality that no individual can survive, leaving large uninhabitable voids that block population spread. Conversely, sites with persistently high environmental quality favor defector dominance, while regions of persistently low environmental quality promote cooperator dominance. As a result, cooperator frequency varies substantially across realizations, even though its average remains close to the noiseless case. Throughout the paper, we use $w = 1$ and $\mu = 0.01$.}
\label{fig2}
\end{figure}

Figure~\ref{fig2} shows representative spatiotemporal population dynamics for $A=0.9$ and $c=0.6$ under different noise conditions. In the noiseless case [Fig.~\ref{fig2}(a)], cooperators and defectors coexist and form spatial domains. In the annealed case [Fig.~\ref{fig2}(b)], the cooperator-occupied region visibly expands relative to the noiseless case. This enhancement can be understood as arising from temporal fluctuations that intermittently create favorable conditions for cooperative expansion. On the other hand, in the quenched case [Fig.~\ref{fig2}(c)], static noise generates spatially localized favorable conditions for either cooperators or defectors, giving rise to distinct frozen spatial domains. These results suggest a shift in the C-dominant and D-dominant phases, and motivate a systematic investigation of the phase boundary between the two phases to quantify the effect of noise.

To characterize the phase behavior analytically, we determine the C-D phase boundary using the MF approach following Ref.\cite{R1}. We examine the competition at the interface between D- and C-clusters to establish the phase boundary. Specifically, we consider a configuration in which a large D-cluster occupies the left and a large C-cluster occupies the right. When an individual at the contact boundary dies, the two clusters separate by some number of empty sites, depending on the local configuration. These vacant sites are subsequently filled as the two clusters expand into the gap from either side. Since each cluster expands at its own habitat growth rate, the fraction of the gap eventually claimed by defectors is $v={V^D}/{(V^D+V^C)}$, where $V^C$ and $V^D$ are the habitat growth rates for C- and D-clusters, respectively. The remaining fraction $1-v$ goes to cooperators. This means that a faster-growing cluster reclaims more vacant space before the next contact event, effectively shifting the boundary in its favor. C-clusters generally have a higher growth rate, while the death probability at the contact boundary is elevated for cooperators. When these two effects are balanced the contact boundary remains stationary on average. 

Taking into account all possible configurations near the contact boundary, we calculate the mean displacement of the D-cluster boundary, $\langle\Delta R_D\rangle$, between successive contact events as
\begin{equation}
    \langle\Delta R_D\rangle= v \left(\frac{\alpha^D_1}{\rho^D}+\frac{\alpha^C_0}{\rho^C} \right)-\frac{\alpha^D_1}{\rho^D}+v(\alpha^C_1-\alpha^C_0)\rho^C~,
    \label{eq:deltaRD}
\end{equation}
where $\alpha^D_1={1}/{(1+Ae^b)}$ is the death probability of the boundary defector facing one cooperating neighbor, and $\alpha^C_0={1}/{(1+Ae^{-c})}$ and $\alpha^C_1={1}/{(1+Ae^{b-2c})}$ are the death probabilities of the boundary cooperator facing zero and one cooperating neighbor, respectively. The densities $\rho_C$ and $\rho_D$ denote the densities of C- and D-clusters, respectively. The C-D phase boundary is obtained by numerically solving $\langle\Delta R_D\rangle=0$ for each value of cost $c$. To this end, we need to estimate $\rho_D$, $\rho_C$, and $v$.

Following Ref.\cite{R1}, we estimate $\rho_D$, $\rho_C$, and $v$ using the MF calculation for homogeneous populations, a pure D-population and a pure C-population, as 
\begin{equation}
    \rho^D\approx\rho^D_\mathrm{MF}=\frac{3-\sqrt{1+4\alpha_0}}{2}~,
    \label{eq:rhoD}
\end{equation}
\begin{equation}
    \rho^C\approx\rho^C_\mathrm{MF}=\frac{3-2\alpha_0+2\alpha_1-\sqrt{(1-2\alpha_1)^2+4(2-\alpha_0)\alpha_2}}{2(1+2\alpha_1-\alpha_0-\alpha_2)}~,
    \label{eq:rhoC}
\end{equation}
and
\begin{equation}
    v\approx v_\mathrm{MF}=\frac{V^D_\mathrm{MF}}{V^C_\mathrm{MF}+V^D_\mathrm{MF}}=\frac{\left(1-\frac{\alpha_0}{\rho^D_\mathrm{MF}}\right)}{\left(1-\frac{\alpha_0}{\rho^D_\mathrm{MF}}\right)+\left(1-\frac{\alpha_0}{\rho^C_\mathrm{MF}}+(\alpha_0-\alpha_1)\rho^C_\mathrm{MF}\right)}~.
    \label{eq:v}
\end{equation}
The death probabilities of a cooperator with $l\in\{0,1,2\}$ cooperating neighbors in a pure C-population are $\alpha_l={1}/{[1+Ae^{l(b-c)}]}$, and the death probability of a defector in a homogeneous D-population equals $\alpha_0$. In the noiseless case, the C-D phase boundary is obtained by solving $\langle\Delta R_D\rangle=0$ with Eqs.~\eqref{eq:rhoD}-\eqref{eq:v}.

We now describe how annealed noise is incorporated into this framework. The environmental quality fundamentally affects death probabilities. Thus, firstly, we examine how the death probabilities are perturbed by environmental noise. Let us consider a smooth function $f(A_i)=f(A+\xi_i)$. For $A\gg \xi_i$, we can assume $f(A_i)\approx f(A)+\xi_i f'(A)+\xi_i^2 f''(A)/2$. As the noise $\xi_i$ is sampled at each MC step, the time average of the function $f(A_i)$ follows $\langle f(A_i)\rangle=\int f(A_i)P(\xi)d\xi$, where $P(\xi)=\exp(-{\xi^2}/{2\sigma^2})/{\sqrt{2\pi\sigma^2}}$ is the Gaussian probability density\cite{R6, R7}. Since the distribution is symmetric about zero, all odd moments vanish so that $\langle\xi_i\rangle=0$, and thus
\begin{equation}
    \langle f(A+\xi_i)\rangle\approx f(A)+\frac{\sigma^2}{2}f''(A).
\end{equation}
Applying the above result, we calculated the average death probabilities with noise,
\begin{equation}
    \langle\alpha^D_1\rangle\approx\alpha^D_1(A)+\sigma^2\frac{e^{2b}}{(1+Ae^b)^3},
\end{equation}
\begin{equation}
    \langle\alpha^C_{0}\rangle\approx\alpha^C_{0}(A)+\sigma^2\frac{e^{-2c}}{(1+Ae^{-c})^3},
\end{equation}
\begin{equation}
    \langle\alpha^C_{1}\rangle\approx\alpha^C_{1}(A)+\sigma^2\frac{e^{2(b-2c)}}{(1+Ae^{b-2c})^3},
\end{equation}
and
\begin{equation}
    \langle\alpha_l\rangle\approx\alpha_l(A)+\sigma^2\frac{e^{2l(b-c)}}{[1+Ae^{l(b-c)}]^3} {{~~\text{ for }}} l=0,1,2~.
\end{equation}
Since the correction term is always positive, noise systematically increases the average death probability for all individual types. However, the magnitude of the correction differs between cooperators and defectors and varies across the $A$-$c$ parameter space, which produces the non-uniform phase boundary shifts.

Substituting the average death probabilities into Eqs.~\eqref{eq:rhoD}-\eqref{eq:v}, we obtain densities and habitat growth rate ratio in the presence of annealed environmental noise:
\begin{equation}
    \tilde{\rho}^D=\frac{3-\sqrt{1+4\langle\alpha_0\rangle}}{2},
    \label{eq:rhoD_noise}
\end{equation}
\begin{equation}
    \tilde{\rho}^C=\frac{3-2\langle\alpha_0\rangle+2\langle\alpha_1\rangle-\sqrt{(1-2\langle\alpha_1\rangle)^2+4(2-\langle\alpha_0\rangle)\langle\alpha_2\rangle}}{2(1+2\langle\alpha_1\rangle-\langle\alpha_0\rangle-\langle\alpha_2\rangle)},
    \label{eq:rhoC_noise}
\end{equation}
and
\begin{equation}
    \tilde{v}=\frac{\tilde{V}^D_\mathrm{MF}}{\tilde{V}^C_\mathrm{MF}+\tilde{V}^D_\mathrm{MF}}=\frac{\left(1-\frac{\langle\alpha_0\rangle}{\tilde{\rho}^D_\mathrm{MF}}\right)}{\left(1-\frac{\langle\alpha_0\rangle}{\tilde{\rho}^D_\mathrm{MF}}\right)+\left(1-\frac{\langle\alpha_0\rangle}{\tilde{\rho}^C_\mathrm{MF}}
    +(\langle\alpha_0\rangle-\langle\alpha_1\rangle)\tilde{\rho}^C_\mathrm{MF}\right)}~.
    \label{eq:v_noise}
\end{equation}
The C-D phase boundary under annealed noise is obtained by solving $\langle\Delta{\tilde{R}}_D\rangle=0$ with Eqs.~\eqref{eq:rhoD_noise}-\eqref{eq:v_noise}. 

For the absorbing boundary under annealed noise, we solve $\tilde{V}^C_\mathrm{MF}=0$. This condition identifies the absorbing boundary because $\tilde{V}^C_\mathrm{MF}=0$ marks the point at which the cooperator habitat ceases to grow. Although this boundary is strictly defined for a pure C-population, it remains a valid approximation for small values of $\mu$, since the absorbing boundary shifts only weakly with mutation rate and stays close to the pure C-population limit.

\begin{figure}[t!]
\includegraphics[width=0.9\linewidth]{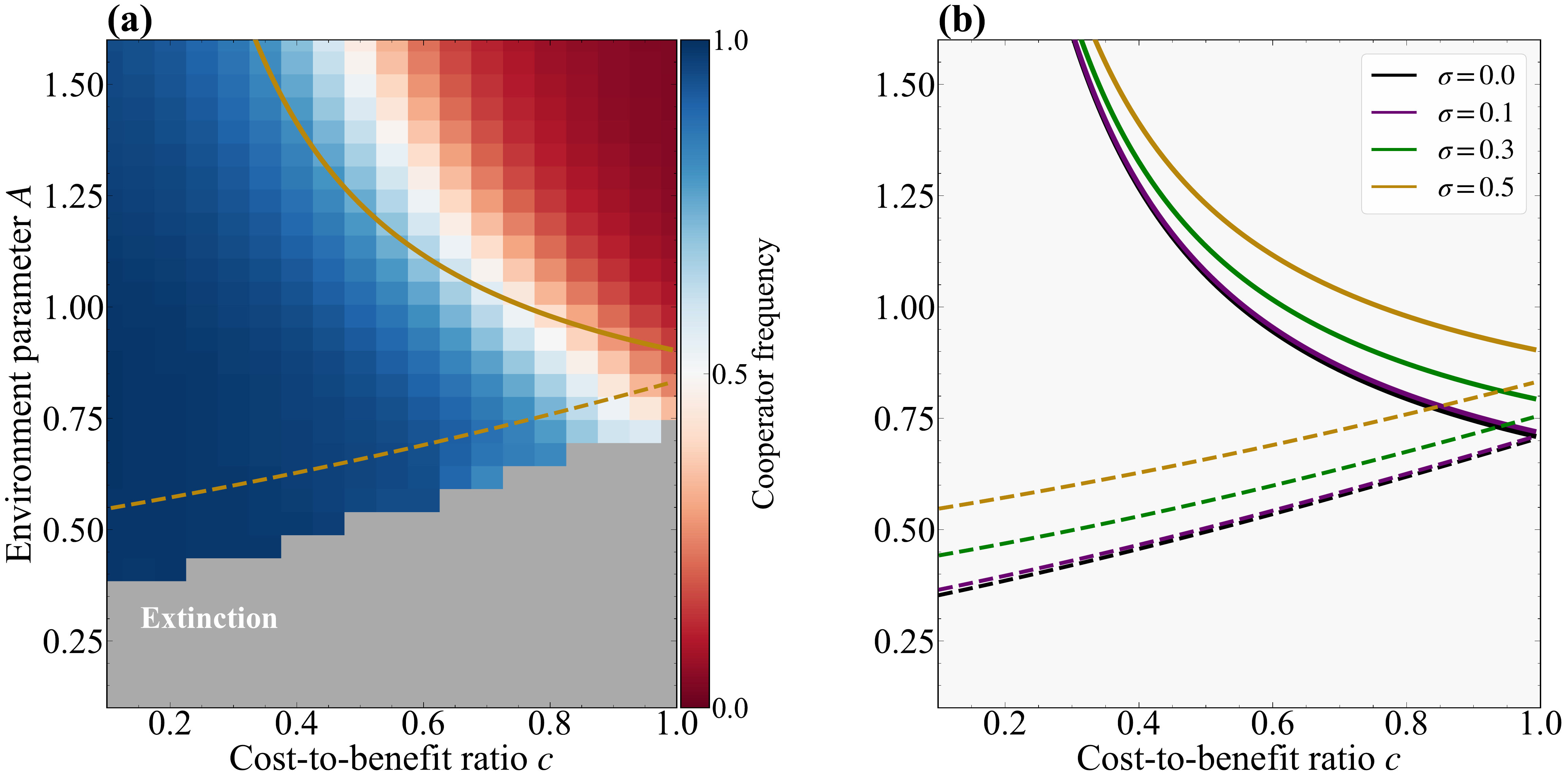}
\caption{ Cooperator frequencies and effect of annealed environmental noise on cooperation evolution. (a) Heat map of the steady-state cooperator frequency $x_C$ in the $A$–$c$ parameter space for $\sigma = 0.5$ (annealed), with the theoretical C-D phase boundary (dark-gold solid curve) and absorbing boundary (dark-gold dashed curve) overlaid. The theoretical boundaries well delineate the extinction (gray region), C-dominant, and D-dominant phases. (b) Theoretical C-D phase boundaries (solid) and absorbing boundaries (dashed) for $\sigma = 0.0, 0.1, 0.3$, and $0.5$, showing that both boundaries shift upward with increasing $\sigma$. Solid and dashed curves denote the theoretical C-D phase boundary and the absorbing boundary, respectively. 
}
\label{fig3}
\end{figure}

\begin{figure}[b!]
\includegraphics[width=0.9\linewidth]{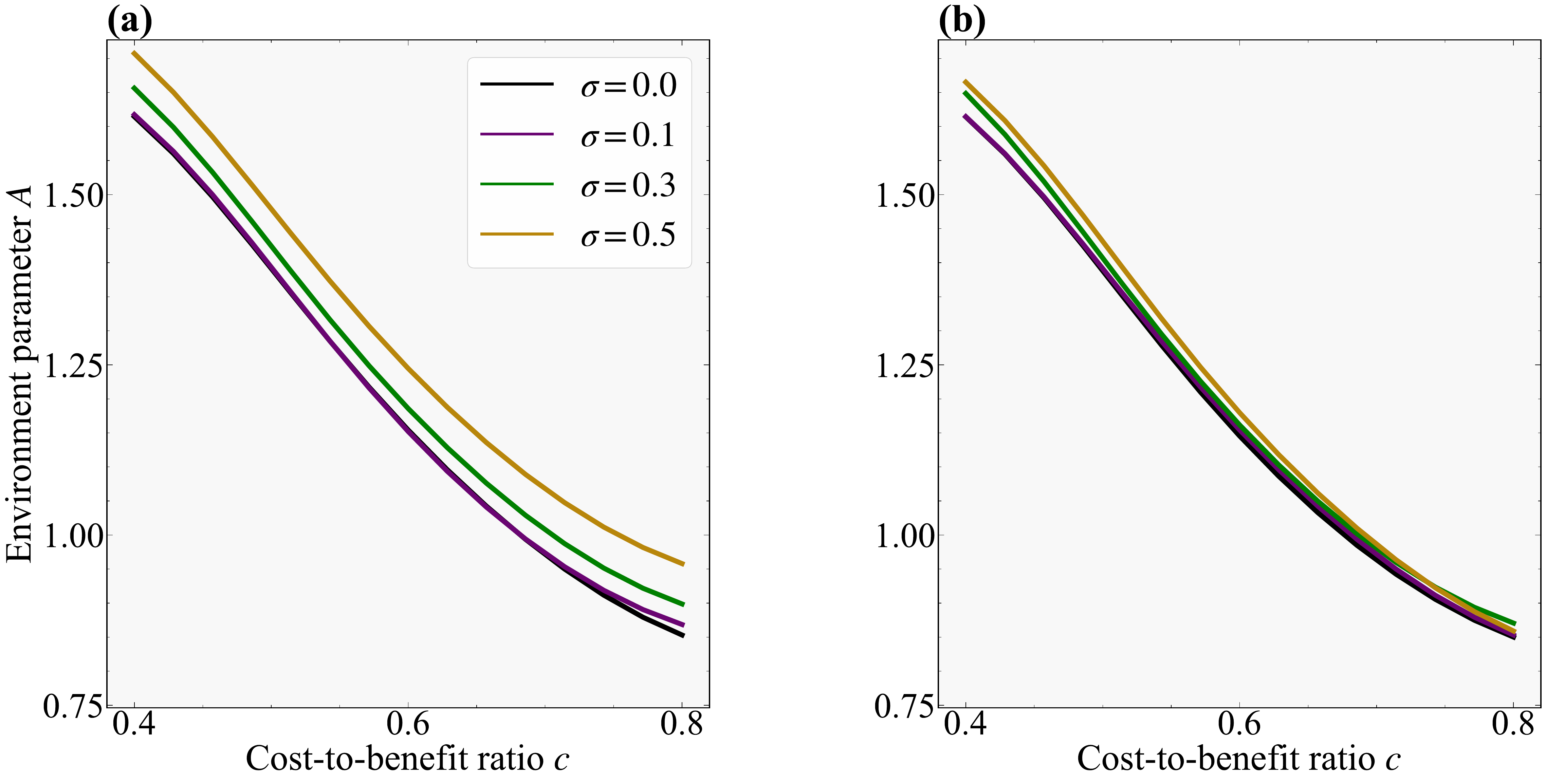}
\caption{Comparison of C-D boundaries for annealed and quenched environmental noise. Numerically obtained C-D boundaries for $\sigma = 0.0, 0.1, 0.3$ and $0.5$ under (a) annealed noise and (b) quenched noise, respectively. In contrast to the annealed case, under quenched noise, the C-D boundaries remain nearly indistinguishable across all values of $\sigma$. We numerically determine the C–D boundaries as the loci where the average cooperator frequency equals $1/2$ in steady state for $M=1000$ ensembles.}
\label{fig4}
\end{figure}

Figure~\ref{fig3} illustrates cooperator frequencies under annealed noise. The extinction and C-D boundaries derived from the MF calculation reasonably well delineate the extinction, C-dominant, and D-dominant phases, as shown in Fig.~\ref{fig3}(a). As $\sigma$ increases, both boundaries shift upward in the $A$-$c$ parameter space [Fig.~\ref{fig3}(b)], revealing two competing effects of annealed noise on spreading cooperation: it broadens the region in which cooperators can invade defectors and persist, while at the same time enlarging the extinction region. The results indicate that spatial environmental noise promotes cooperation, provided the average environmental quality is not too low.

We also compute the C-D phase boundaries numerically, and the results are shown in Fig.\ref{fig4}. For the annealed case, C-D phase boundaries shift upward by progressively larger amounts with $\sigma$: the separation between the $\sigma=0.3$ and $\sigma=0.5$ boundaries is greater than that between the $\sigma=0.1$ and $\sigma=0.3$ boundaries. This accelerating, non-uniform shift is a direct consequence of the $\sigma^2$ dependence of the noise-averaged death probabilities, and it is faithfully reproduced by the MF prediction [Fig.~\ref{fig3}(b)]. The agreement demonstrates that the MF theory captures not merely the direction of the boundary shift but also this qualitative feature of how the spacing between successive boundaries grows with noise intensity.

In contrast to the annealed case, under quenched noise, a closed-form MF theory is not available. We therefore obtain only numerical C-D phase boundaries [see Fig.~\ref{fig4}(b)]. For quenched noise, the C-D boundaries remain nearly indistinguishable across all values of $\sigma$, indicating that static spatial heterogeneity alone has only a weak effect on the phase boundary location. This suggests that temporal fluctuations, rather than static spatial disorder, are the primary driver of noise-induced shifts in the cooperative phase structure.
\section{CONCLUSIONS}
In this work, we have investigated the effects of two distinct types of spatial environmental noise, annealed and quenched, on the evolution of cooperation in a one-dimensional spatial evolutionary game model\cite{R1, R3}. 
For annealed noise, in which the environmental quality fluctuates independently at each site and each time step\cite{R4}, we developed a mean-field (MF) theory by replacing the noise-dependent death probabilities with their distribution-averaged values\cite{R6, R7}. The key analytical result is that noise systematically increases the effective death probability for all individual types, since the second-order correction is strictly positive. This leads to an upward shift of both the cooperator-defector (C-D) phase boundary and the absorbing boundary in the $A$-$c$ parameter space as $\sigma$ increases. Consequently, annealed noise can promote cooperation when cooperators and defectors are most evenly matched in the absence of noise. In that regime, increasing the noise intensity $\sigma$ progressively tips the competition in favor of cooperators, so that parameter regions that are defector-dominant in the noiseless case become cooperator-dominant once the annealed noise is sufficiently strong. Our MF framework, validated against simulations, provides a tractable analytical approach for understanding how environmental stochasticity shapes the evolution of cooperation in spatially structured populations\cite{R1, R4, R6, R7}.

On the other hand, for quenched noise, in which each site is assigned a permanently fixed random environmental value representing persistent spatial heterogeneity\cite{R5}, the C-D phase boundary remains nearly unchanged across all noise levels, indicating that static spatial heterogeneity has little effect on the domain boundary dynamics. We attribute this null result to the spatially uncorrelated nature of the disorder: because an independent value is drawn at every site, the noise is too fine-grained for any extended favorable or unfavorable region to form. This is why the quenched case so closely resembles the noiseless one.

A natural direction for future work follows directly from this interpretation. Our quenched noise is uncorrelated from site to site, which represents the limit most strongly washed out by spatial averaging. A more realistic environment, however, is often correlated over some characteristic length scale, so that neighboring sites share similar conditions and coherent patches of favorable or unfavorable habitat can persist\cite{R5, R8}. Introducing spatially correlated quenched noise with a tunable correlation length would interpolate between the uncorrelated limit studied here and a fully homogeneous shift of the environment, and could reveal whether extended static patches, unlike fine-grained disorder, can relocate the C-D phase boundary. Exploring how the correlation length of static environmental heterogeneity competes with the temporal fluctuations of annealed noise is, in our view, a promising avenue for clarifying the roles of space and time in the evolution of cooperation.


\begin{acknowledgments}
This work was supported by INHA UNIVERSITY Research Grant.
\end{acknowledgments}



\end{document}